\documentclass[twocolumn,nopacs,preprintnumbers,amsmath,amssymb]{revtex4}

\usepackage{verbatim}
\usepackage{graphicx}
\usepackage{dcolumn}
\usepackage{bm}

\begin{document}

\title{Enhanced field emission from multiwall carbon nanotube films by secondary growth}
\author{Christian Klinke}
\email{klinke@chemie.uni-hamburg.de}
\affiliation{Institut de Physique des Nanostructures, Ecole Polytechnique Federale de Lausanne, CH-1015 Lausanne, Switzerland}
\affiliation{IBM T. J. Watson Research Center, Yorktown Heights, NY 10598, USA}
\author{Erik Delvigne}
\affiliation{Institut de Physique des Nanostructures, Ecole Polytechnique Federale de Lausanne, CH-1015 Lausanne, Switzerland}
\author{Johannes V. Barth}
\affiliation{Institut de Physique des Nanostructures, Ecole Polytechnique Federale de Lausanne, CH-1015 Lausanne, Switzerland}
\affiliation{Department of Chemistry and Physics \& Astronomy, University of British Columbia, Vancouver, Canada}
\author{Klaus Kern}
\affiliation{Institut de Physique des Nanostructures, Ecole Polytechnique Federale de Lausanne, CH-1015 Lausanne, Switzerland}
\affiliation{Max-Planck Institute for Solid State Research, D-70569 Stuttgart, Germany}

\begin{abstract}

We have studied nickel, gold, and ferritin coatings on catalytically grown multiwall carbon nanotubes, as well as the generation of secondary nanotubes by resubmitting the decorated nanotubes to the chemical vapor deposition process. Nickel layers sputtered on nanotubes show a stronger interaction with the nanotube walls than gold coatings. At ambient temperature this results in a metal film that is more homogeneous for Ni than for Au. Surface mass transport at elevated temperatures leads to a transformation of the coating to nanoscale clusters on the nanotube surface. The resulting Au clusters are sphere-like with a very small contact area with the nanotube whereas the Ni clusters are stretched along the tube axis and have a large contact area. Secondary nanotubes were established by growing nanotubes directly on the walls of primary nanotubes. Thin Ni layers or ferritin served as catalyst. We compared the field emission properties of samples with and without secondary nanotubes. The presence of secondary nanotubes enhances the field emission substantially.

\end{abstract}

\maketitle

\section{Introduction}

Due to their exceptional mechanical and electrical properties, carbon nanotubes (CNT) have been studied intensively in the last decade, and are now considered for application in real devices. For example, they can be used as single-molecule transistors [1,2], or as interconnects on chips [3]. Other applications of nanotubes include sensors [4], emitters for light sources based on field emission [5,6], and electron sources in transmission electron microscopes [7]. Recently, the first working flat panel display based on carbon nanotube field emitters has been reported [8]. Nanotubes can serve as well as a conductive backbone for chains of metal clusters [9,10]. These clusters may be used as electrodes or as catalytic particles with a large surface area. Coating nanotubes with ferromagnetic materials can provide very precise tips for magnetic force microscopy (MFM) and spin-polarized scanning tunneling microscopy, and thus improve the performance of such methods. Cross-linking of carbon nanotubes can be important for the creation of nanotube networks for microelectronics. Moreover, if the cross-links between the nanotubes establish a stable connection, it should also be possible to create very strong polymer compound materials. Up to now the mechanical strength of nanotube-polymer composites is much lower than that of pure nanotubes [11] because the nanotubes in compound materials slip out of the polymer matrix when large forces are applied. Nanotube networks made by cross-linking can, in principle, alleviate this effect.

In this contribution we show that a second stage of nanotubes on top of primary nanotube films can enhance the field emission. Such films potentially can improve the homogeneity and efficiency of field emission flat panel displays. In order to reach this goal we studied the adsorption of the catalyst metal nickel on multiwall carbon nanotubes, and compared the results to the behavior of more inert gold coatings. The nickel coating was then used to grow a second generation of nanotubes on the walls of the primary nanotubes. We also successfully employed ferritin to generate secondary tubes which resulted in secondary nanotubes of higher uniformity. This kind of nanotube network could be further cross-linked by tertiary connections (nanotubes on nanotubes on nanotubes) and so forth. Such networks might help to improve nanotube compound materials. The two configurations (with and without secondary nanotubes) have been compared with respect to their field emission properties. Secondary nanotubes enhance the extracted current and the homogeneity of field emission substantially.

\begin{figure}[ht]
\begin{center}
\includegraphics[width=0.45\textwidth]{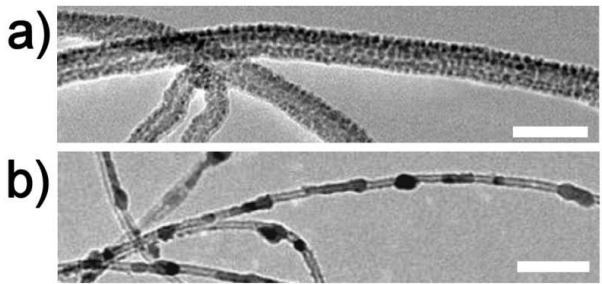}
\caption {\it TEM micrographs of multiwall CNTs with a nickel coating. a) Nominal 5~nm nickel layer sputtered at 20$^{\circ}$C, showing small clusters with a high nucleation density. The scale bar corresponds to 40~nm. b) Nickel layer upon heating to 660$^{\circ}$C for 2~min. Larger clusters coexist with segments of continuous coating on the nanotube. The scale bar corresponds to 100~nm.}
\end{center}
\end{figure}

\section{Experimental Methods}

\textbf{Generation of primary nanotubes.} We produced the carbon nanotubes by using either 500~$\mu$g/ml ferritin + 50~mMol Al(NO$_{3}$)$_{3}$ $\cdot$ 9H$_{2}$O in demineralized water [12] as a catalyst which yielded in thin multiwall nanotubes ($\sim$~5~nm) or a 50~mM ferric nitrate solution [13] which produced tubes with a diameter between 10 and 20~nm. All samples were submitted for 5~min to a chemical vapor deposition (CVD) process using 20~mbar C$_{2}$H$_{2}$ at 660$^{\circ}$C directly after deposition of the catalyst material. As substrates we used tungsten TEM grids ($\oslash$ 3~mm) for the structural investigations and silicon dies (1x1~cm$^{2}$) for the field emission experiments. In the latter case the catalyst was delivered to the substrate by microcontact printing.

\textbf{Generation of secondary nanotubes.} For the growth of secondary nanotubes, either nickel was sputtered on the primary nanotubes, or the samples have been dipped into the ferritin solution and dried in air. The samples were then resubmitted to the CVD process.

\textbf{Transmission electron microscopy.} For the characterization of the metal layers on the nanotubes and the in-situ observation of their behavior under heating we introduced the substrate grids into a TEM (Philips EM 430 ST, operating at 300~kV) using a sample holder which is resistively heatable up to 1000$^{\circ}$C. 

\textbf{Field emission measurements.} The field emission measurements were performed using the nanotube samples as cathodes. The emitted electrons were collected on a highly polished stainless steel spherical counterelectrode of 1~cm diameter, which corresponds to an emission area of $\sim$~0.007~cm$^{2}$ [6]. The distance between the electrodes was adjusted to 125~$\mu$m. A Keithley 237 source-measure unit was used to supply the voltage (up to 1000~V) and to measure the current with pA sensitivity, allowing the characterization of current-voltage (I-V) behavior.

\begin{figure}[ht]
\begin{center}
\includegraphics[width=0.45\textwidth]{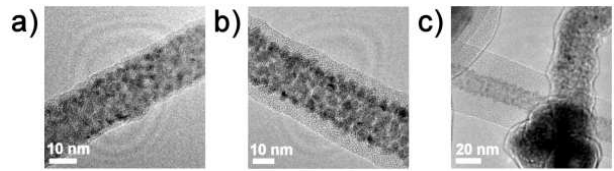}
\caption {\it TEM micrographs of ferric nitrate catalyzed CNTs with nominal a) 5~nm, b) 10~nm and c) 15~nm nickel coating. A cluster-like nickel under-layer is visible, with a more electron transparent continuous nickel top-layer. The electron transparency is especially in c) clearly observable.}
\end{center}
\end{figure}

\section{Results and discussion}

In order to realize higher field emission currents at given extraction voltages and better areal homogeneity of emitters by a second stage of nanotubes, we first analyzed the behavior of nickel decorated nanotubes at elevated temperatures.

\textbf{Wetting behavior of metal on nanotubes.} We grew ferritin catalyzed CNTs on TEM grids and metallized them with nickel. The nominal 5~nm Ni layer of Fig.~1a shows a cluster-like structure with high nucleation density, leaving almost no parts of the nanotube uncovered. Metal deposited on graphite or on nanotubes usually tend to yield in a low nucleation density which is caused by the weak condensate-substrate interaction [14]. Nickel has a relatively strong interaction with the graphite-like nanotube surface, which is attributed to curvature-induced rehybridization of carbon sp$^{2}$ orbitals with the Ni d-orbital [15]. The Ni coatings undergo a shape transition upon heating, as can be seen in Fig.~1b after heating to 660$^{\circ}$C for 2~min. Upon heating the clusters become bigger and the number of clusters per surface area is greatly reduced. Parts of the nanotubes are covered by segments of continuous coating, which did not exist beforehand. 

Fig.~2 shows nanotubes covered with nominal 5, 10 and 15~nm Ni. A cluster-like underlayer, as aforementioned, was observed in all samples. During an initial phase Ni clusters are formed on the nanotube, whereas additional Ni forms a continuous layer on top of the clusters. The amount of deposited Ni only influences the thickness of the continuous layer and does not affect the cluster-like underlayer. An interesting observation is the difference in contrast between the dark cluster-like underlayer and the more electron transparent top layer. After comparison with TEM images of Ni in other publications, it could be that the dark imaged parts are crystalline Ni [16,17,18] and the more electron transparent parts are amorphous Ni [19,20]. However, it is unclear what could cause such a change from crystalline to amorphous Ni during film growth. 

In order to classify the behavior of the nickel/nanotube system, we compare it with nanotubes decorated with nominal 5~nm Au. Gold aggregates on the nanotube surface as a discontinuous pattern of small nanoscopic islands (see Fig.~3a) that presumably first decorate defects at the nanotube surface [21]. The Au film has a low nucleation density, leaving parts of the nanotube between the islands uncovered. Elevated temperatures causes the small Au clusters to merge into isolated large particles due to the increased mobility and mass transport [22,23]. The temperature dependent changes of the Au coating have been studied by in-situ heating in the TEM. Upon heating the Au, islands initially change in shape and form more sphere-like clusters in order to minimize the surface energy. Further heating to higher temperatures leads to integration of smaller clusters into bigger ones by diffusion of Au on the nanotube surface, known as Ostwald ripening [24]. This reduces the number of clusters and the remaining clusters grow in diameter, the result after heating to 660$^{\circ}$C for 2~min is shown in Fig.~3b. At temperatures over 1000$^{\circ}$C the Au clusters start to disappear which is associated with the evaporation of the Au clusters due to the high vapor pressure of those objects at elevated temperatures. The macroscopic vapor pressure of Au is about 10-5~mbar at 1024$^{\circ}$C [25], while the base pressure in a TEM is clearly below this vapor pressure. Gold decorated multiwall nanotubes could be used to grow silicon nanowires on the sidewall of the tubes. This would render a interesting semiconductor-metal junction.

\begin{figure}[ht]
\begin{center}
\includegraphics[width=0.45\textwidth]{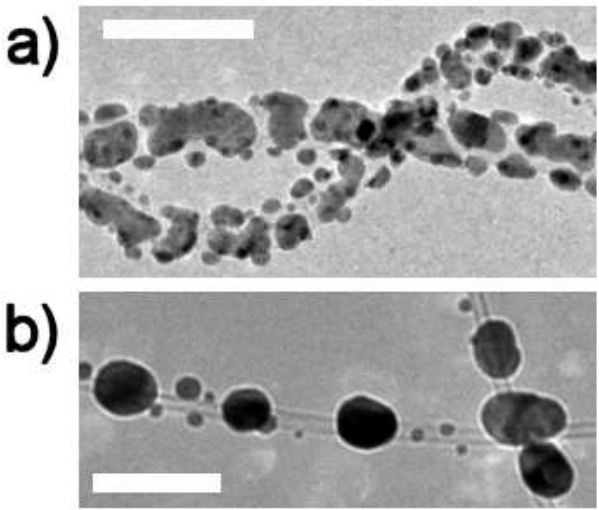}
\caption {\it TEM micrographs of multiwall CNTs with a gold coating. a) Nominal 5~nm gold layer sputtered at 20$^{\circ}$C, showing gold islands with an arbitrary shape. b) Same Au layer after heating to 660$^{\circ}$C for 2~min, showing the formation of sphere-like clusters which have a small contact area with the nanotubes. Scale bars correspond to 50~nm.}
\end{center}
\end{figure}

\textbf{Growth of secondary nanotubes.} Iron, cobalt and nickel are known to be good catalysts for the growth of carbon nanotubes by CVD [13]. Even thin layers of these metals can be used for growing nanotubes because the thin layer will fragmentize into small clusters when heated, as demonstrated above. We used Ni coated carbon nanotubes to grow secondary carbon nanotubes on the sidewalls of the primary carbon nanotubes. A nominal 5~nm Ni layer was sputtered onto ferric nitrate catalyzed primary nanotubes (Fig.~4a) and onto thinner ferritin catalyzed primary nanotubes (Fig.~4b). The samples were then re-submitted to the same standard CVD process for 5~min. As a result, secondary nanotubes were obtained on the sidewalls of the primary nanotubes. For both types of primary nanotubes the nickel-catalyzed secondary nanotubes have a diameter of about 10~nm. In addition to the secondary nanotubes, we found also Ni clusters that did not result in the growth of secondary nanotubes. These are, on the one hand, the segments of the continuous Ni layer with a comparatively large contact area, and on the other hand Ni clusters that did not acquire the correct dimensions required for nanotube growth (too small or too big), or that were already poisoned by an excess of amorphous carbon before nanotube growth could effectively set in. The secondary nanotubes are less graphitized and shorter in length than the primary nanotubes. The reduced length is possibly caused by a faster poisoning of the catalyst particles by amorphous carbon on the catalytically active surfaces. 

In order to enhance the uniformity of the secondary nanotubes we used ferritin as a catalyst. After the growth of primary ferric nitrate catalyzed CNTs on TEM grids, the samples have been dipped into a ferritin solution, dried, and finally re-submitted to the same CVD process. This method provides the advantage of a very homogenous metal cluster size which is defined by the ferritin core [12]. Indeed, thin nanotubes were found on the sidewalls of the thicker primary nanotubes, as demonstrated in Fig.~4c. Again the secondary nanotubes are relatively short and apparently less graphitized than the thicker and longer primary nanotubes. In contrast to the Ni cluster case, all the catalyst particles that are present on the primary nanotubes resulted in the growth of secondary nanotubes. From this observation we conclude that the ferritin cores are very effectively transformed into catalyst particles that have the correct dimensions for nanotube growth. It is worth noting that the catalyst particles are always located at the tops of the secondary nanotubes (both nickel and ferritin catalyzed), i.e. all secondary nanotubes grow via the top-growth mechanism, whereas ferritin nanotubes grown on Si/SiO$^{2}$ substrates usually grow via the base-growth mechanism [26]. We attribute this to the weak interactions between the metal catalyst particles and the graphite-like surface of the nanotubes.

\begin{figure}[ht]
\begin{center}
\includegraphics[width=0.45\textwidth]{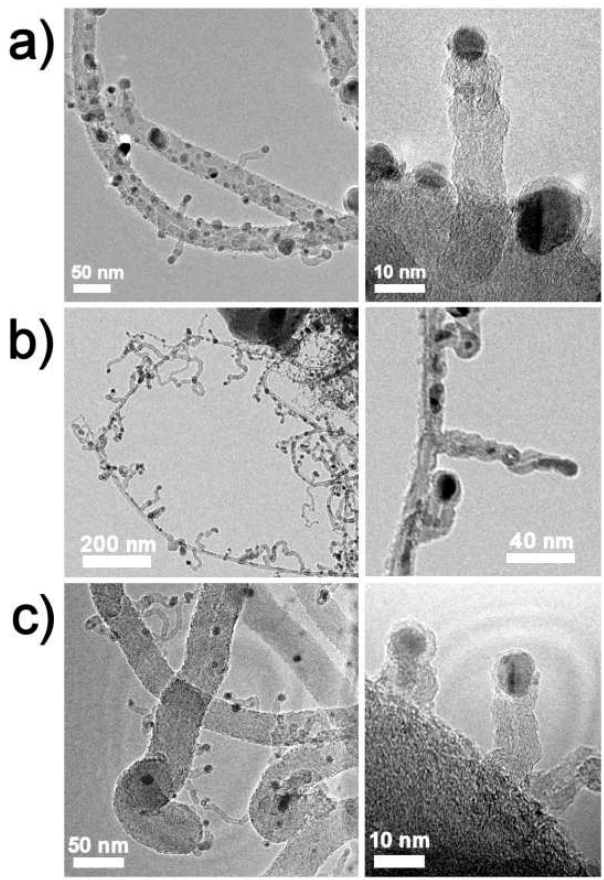}
\caption {\it TEM micrographs of secondary CNTs grown on the sidewalls of primary CNTs. a) Secondary nickel catalyzed CNTs grown on relatively thick primary ferric nitrate catalyzed CNTs. b) Secondary nickel catalyzed CNTs grown on thin primary ferritin catalyzed CNTs. c) Secondary ferritin catalyzed CNTs grown on primary ferric nitrate catalyzed CNTs.}
\end{center}
\end{figure}

\textbf{Field emission with secondary carbon nanotubes.} The influence of the secondary nanotubes on the field emission properties of nanotube samples has been investigated. A sample with ferritin catalyzed CNTs on a Si/SiO$_{2}$ substrate without secondary nanotubes (black dashed line in Fig.~5) has a turn-on field E$_{to}$ of 4.8~V/$\mu$m (field to obtain a current density of 10-5~A/cm$^{2}$; first illumination of a screen pixel) and a threshold field E$_{th}$ of 7.9~V/$\mu$m (field resulting in a current density of 10-2~A/cm$^{2}$; saturation of a screen pixel). SEM observations show that the nanotubes grow in random direction and thus, an arbitrary part of the wall of the individual nanotubes (opposed to the tip) points towards the counter electrode. This has two negative effects for the emitted current density: firstly this reduces the field amplification factor for most of the nanotubes, secondly, since only the nanotubes with highest field amplification will emit, the actual emitter density will probably be low in spite of the high nanotube density [27]. When nickel catalyzed secondary nanotubes are grown (black solid line), the field emission properties improve (i.e. a larger emitted current density for a certain applied electric field), with now E$_{th}$ = 6.6~V/$\mu$m. The secondary nanotubes are not significantly thinner than the primary ferritin catalyzed CNTs (see also Fig.~4b) and therefore the improvement cannot be ascribed to a smaller radius of curvature at the tip. In fact, the deduced field amplification is even about 10~\% smaller, possibly because of the short length of the secondary nanotubes. We assume that the field emission is improved by secondary nanotubes that are located at the sidewall parts of the primary nanotubes that are close to the counter electrode. These parts of the primary tubes would normally not emit electrons because there are not the nanotube tips. The secondary nanotubes located there facilitate field emission from an increased number of emitters per surface area. 

A similar experiment was conducted for the thicker ferric nitrate catalyzed primary CNTs. The sample with just primary CNTs (dashed gray line) has a turn-on field E$_{to}$ = 5.2~V/$\mu$m. A threshold field was not achieved and must be above 8~V/$\mu$m. These field emission characteristics are inferior to the ferritin catalyzed CNT sample, most probably caused by the larger diameter of the nanotubes. The growth of secondary nickel catalyzed CNTs (solid gray line) results in a large improvement in the field emission (E$_{to}$ = 5.0~V/$\mu$m and E$_{th}$ = 6.7~V/$\mu$m). We found that there is no difference in the field amplification factors between the two samples. The negative influence of the short length of the secondary nanotubes is balanced by the positive influence of the smaller radius of curvature of the secondary nanotubes. The improved field emission is, as for the ferritin catalyzed CNTs associated with the increased number of emitters. In fact the field emission curves of the two samples with secondary nanotubes in Fig.~5 are similar. Apparently the diameter of the primary nanotubes has very little influence on the field emission when secondary nanotubes are present. 

\begin{figure}[ht]
\begin{center}
\includegraphics[width=0.45\textwidth]{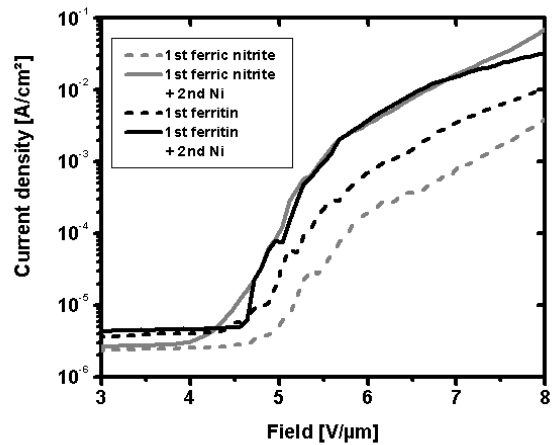}
\caption {\it Field emission measurements on nanotube samples. Comparison of the field emission obtained for samples with ferritin catalyzed primary CNTs (black lines), with and without nickel catalyzed secondary CNTs. Idem for samples with ferric nitrate catalyzed primary CNTs (gray lines), with and without nickel catalyzed secondary CNTs.}
\end{center}
\end{figure}

The reproducibility of the field emission results using nickel as catalyst for the regrowth is very good. None of the samples with nickel catalyzed regrowth shows any major deviation from the curves shown in Fig.~5. Using Ferritin as catalyst we could reach a quite homogeneous distribution of secondary nanotubes on the used TEM grids. Those samples are smaller than the silicon dies and provide sparse distribution of nanotubes on the substrate. On the silicon chips we could not reach a uniform distribution of ferritin on primary nanotubes. The reason for this might lie in the dense film of primary nanotubes and the hydrophobicity of the graphite-like nanotubes. The uneven distribution of ferritin resulted in an unclear picture of the field emission of ferritin-catalyzed secondary nanotubes.

The fact that the use of secondary nanotubes on the sidewalls of carbon nanotube films enhances the field emission substantially has great implications on the production of field emission devices like flat panel displays. It can increase the efficiency, the homogeneity, and the reliability of according devices. 

\section{Conclusion}

Ni and Au coatings sputtered on nanotubes have been studied at different temperatures. Ni shows a stronger interaction with the nanotube walls than Au. This results in a coating at ambient temperature that is more homogeneous for Ni than for Au. Heating to 660$^{\circ}$C leads to the formation of larger clusters on the nanotube surface. The Au clusters are sphere-like with a very small contact area with the nanotube, whereas the Ni clusters are stretched along the nanotube direction and have a larger contact area. 

Secondary nanotubes have been established by growing nanotubes directly on the walls of primary nanotubes. Thin Ni layers or ferritin served as catalyst. These secondary nanotubes however attain lengths of generally only 50 to 100~nm. The presence of secondary nanotubes clearly improves the field emission characteristics of the nanotube samples. The field enhancement factors of the samples do not improve and the improvement in field emission is therefore attributed to an increased number of effective emitters.

\textbf{Acknowledgement.} The Swiss National Science Foundation (SNF) is acknowledged for the financial support. The electron microscopy was performed at the Centre Interdepartmental de Microscopie Electronique (CIME) of the EPFL.

\section*{References}

[1] Tans, S. J.; Verschueren, R. M.; Dekker, C.  Nature 1998, 393, 49.

[2] Martel, R.; Schmidt, T.; Shea, H. R.; Hertel, T.; Avouris, P.; Appl. Phys. Lett. 1998, 73, 2447.

[3] Kreupl, F.; Graham, A. P.; Duesberg, G. S.; Steinhögl, W.; Liebau, M.; Unger, E.; Hoenlein, W. Microelect. Eng. 2002, 64, 399.

[4] Kong, J.; Franklin, N. R.; Zhou, C. W.; Chapline, M. G.; Peng, S.; Cho, K. J.; Dai, H. J. Science 2000, 287, 622.

[5] Saito, Y.; Uemura, S.; Hamaguchi, K. Jpn. J. Appl. Phys. 1998, 37, 346.

[6] Bonard, J. M.; Croci, M.; Klinke, C.; Kurt, R.; Noury, O.; Weiss, N. Carbon 2002, 40, 1715.

[7] de Jonge, N.; Lamy, Y.; Schoots, K.; Oosterkamp, T. H. Nature 2002, 420, 393.

[8] Wang, Q. H.; Setlur, A. A.; Lauerhaas, J. M.; Dai, J. Y.; Seeling, E. W.; Chang, R. P. H. Appl. Phys. Lett. 1998, 72, 2912.

[9] Burghard, M.; Krstic, V.; Duesberg, G. S.; Philipp, G.; Muster, J.; Roth, S.; Journet, C.; Bernier, P. Synth. Metals 1999, 103, 2540.

[10] Banerjee, S.; Wong, S. S. Nano Lett. 2002, 2, 195.

[11] Vigolo, B.; Poulin, P.; Lucas, M.; Launois, P.; Bernier, P. Appl. Phys. Lett. 2002, 81, 1210.

[12] Bonard, J. M.; Chauvin, P.; Klinke, C. Nano Lett. 2002, 2, 665.

[13] Klinke, C.; Bonard, J. M.; Kern, K. Surf. Sci. 2001, 492, 195.

[14] Zhang, Y.; Franklin, N. W.; Chen, R. J.; Dai, H. Chem. Phys. Lett. 2000, 331, 35.

[15] Menon, M.; Andriotis, A. N.; Froudakis, G. E. Chem. Phys. Lett. 2000, 320, 425.

[16] Gao, X. P.; Zhang, Y.; Chen, X.; Pan, G. L.; Yan, J.; Wu, F.; Yuan, H. T.; Song, D. Y. Carbon 2004, 42, 47.

[17] Rojas, T. C.; Sayague, M. J.; Caballero, A.; Koltypin, Y.; Gedanken, A.; Ponsonnet, L.; Vacher, B.; Martin, J. M.; Fernandez, A. J. Mater. Chem. 2000, 10, 715.

[18] Zhang, P.; Zuo, F.; Urban, F. K.; Khabari, A.; Griffiths, P.; Hosseini-Tehrani, A. J. Magn. Magn. Mater. 2001, 225, 337.

[19] McFadden, S. X.; Zhilyaev, A. P.; Mishra, R. S.; Mukherjee, A. K. Mater. Lett. 2000, 45, 345.

[20] Strzeciwilk, D.; Wokulski, Z.; Tkacz, P. Cryst. Res. Technol. 2003, 38, 283.

[21] Fan, Y.; Burghard, M.; Kern, K. Adv. Mat. 2002, 14, 130.

[22] Barfotti, L.; Jensen, P.; Hoareau, A.; Treilleux, M.; Cabaud, B.; Perez, A.; Aires, F. Surf. Sci. 1996, 367, 276.

[23] Klinke, C.; Bonard, J. M.; Kern, K. J. Phys. Chem. B 2004, 108, 11357.

[24] Ostwald, W. Z. Phys. Chem. 1900, 34, 495.

[25] American Institute of Physics handbook. 3rd ed, ed. D.E. Gray. 1972, New York: McGraw-Hill Book Company.

[26] Klinke, C.; Kurt, R.; Bonard, J. M. J. Phys. Chem. B 2002, 106, 11191.

[27] Bonard, J. M.; Weiss, N.; Kind, H.; Stöckli, T.; Forro, L.; Kern, K.; Chatelain, A. Adv. Mater. 2001, 13, 184.

\end{document}